\documentclass[pre,reprint,twocolumn,showpacs,showkeys,pdflatex]{revtex4}
\usepackage{graphicx}
\usepackage{amsfonts}
\usepackage{amsmath}
\usepackage{amssymb}
\usepackage{epstopdf}
\usepackage{color}
\usepackage{bm} 
\newenvironment{nalign}{
    \begin{equation}
    \begin{aligned}
}{
    \end{aligned}
    \end{equation}
    \ignorespacesafterend
}
\begin{document}
%%%%%%%%%%%%%%%%%%%%%%%%%%%%%%%%%%%%%%%%%%%%%%%%%%%%%%%%%%%%%%%%%%%%%%%%%%%%%%%%
\title{Surge of Power Transmission in Flat and Nearly Flat Band Lattices}
%%%%%%%%%%%%%%%%%%%%%%%%%%%%%%%%%%%%%%%%%%%%%%%%%%%%%%%%%%%%%%%%%%%%%%%%%%%%%%%%
\author{H.\ Susanto}
\email{Corresponding author: hadi.susanto@ku.ac.ae}
\affiliation{Department of Mathematics, Khalifa University of Science $\&$ Technology, P.O.\ Box 127788, 
             Abu Dhabi, United Arab Emirates}

\author{N.\ Lazarides}
\affiliation{Department of Mathematics, Khalifa University of Science $\&$ Technology, P.O.\ Box 127788, 
             Abu Dhabi, United Arab Emirates}

\author{I.\ Kourakis}
\affiliation{Department of Mathematics, Khalifa University of Science $\&$ Technology, P.O.\ Box 127788, 
             Abu Dhabi, United Arab Emirates}
%%%%%%%%%%%%%%%%%%%%%%%%%%%%%%%%%%%%%%%%%%%%%%%%%%%%%%%%%%%%%%%%%%%%%%%%%%%%%%%%
\date{\today}
%%%%%%%%%%%%%%%%%%%%%%%%%%%%%%%%%%%%%%%%%%%%%%%%%%%%%%%%%%%%%%%%%%%%%%%%%%%%%%%%
\begin{abstract}
Flat band systems can yield interesting phenomena due to their completely flat band, such as dispersion suppression of waves with frequency at the band. While linear transport vanishes, the corresponding nonlinear case is still an open question. Here, we study power transmission along nonlinear sawtooth lattices due to waves with the flat band frequency injected at one end. While there is no power transfer %from the boundary to the array 
for small intensity, there is a threshold amplitude above which a surge of power transmission occurs, i.e., supratransmission, for defocusing nonlinearity. This is due to a nonlinear evanescent wave with the flat band frequency that becomes unstable. We show that %the novel properties of 
dispersion suppression and supratransmission also exist even when the band is nearly flat. %persist in nearly flat band lattices due to interplay between flat band characteristics and nonlinearity.
\end{abstract}
%%%%%%%%%%%%%%%%%%%%%%%%%%%%%%%%%%%%%%%%%%%%%%%%%%%%%%%%%%%%%%%%%%%%%%%%%%%%%%%%
\maketitle
%%%%%%%%%%%%%%%%%%%%%%%%%%%%%%%%%%%%%%%%%%%%%%%%%%%%%%%%%%%%%%%%%%%%%%%%%%%%%%%%
{\em Introduction.} 
Flat band lattices are periodic media with at least one of their Bloch bands completely flat in the entire Brillouin zone (see, e.g.
\cite{sutherland1986localization,lieb1989two,arai1988strictly} for early theoretical predictions). Because the wave group velocity vanishes, the media are dispersionless and 
as such, wave transport is suppressed and particles will have infinite effective mass, which brought forth exotic characteristics and dynamics (see, e.g., the recent reviews \cite{tang2020photonic,Leykam2018,Leykam2018b,Vicencio2021}).
As the kinetic energy is completely quenched, particle interaction becomes enhanced that makes flat band lattices a perfect candidate to study complex many-body quantum 
states and strongly correlated physical systems \cite{lieb1989two,liu2012fractional,tang2011high}. Linear flat band lattices also support localized modes without inhomonegeity that 
have been demonstrated theoretically \cite{Lopez-Gonzalez2016,maimaiti2017compact,Lazarides2017,Lazarides2019,He2021} and observed experimentally \cite{Vicencio2015,mukherjee2015observation,mukherjee2015observation2,Biesenthal2019}. In interacting with nonlinearity, flat bands can support compactons \cite{lopez2016linear} 
and interaction-driven topological states \cite{di2016topological}.

Here we are interested in the scattering of plane waves by a nonlinear medium which possesses a flat band. Due to the dispersionless property of the medium, what happens when the plane wave frequency falls within the flat band is a 
mystery. 
%is in a nearly-flat or even at the completely flat band.  
One may expect that the wave will be totally reflected, but it is not certain. Moreover, the presence of nonlinearity can alter expected behaviors and the dynamics can become 
complicated. An important example is the so-called supratransmission, i.e., an abrupt transmission phenomenon of plane wave power with frequency in the forbidden band 
when the incoming wave amplitude is above a threshold, that has been reported both theoretically and experimentally 
\cite{taverner1998nonlinear,geniet2002energy,khomeriki2004nonlinear,susanto2008boundary,susanto2008calculated,koon2014experimental}.

Here, we report a novel supratransmission of plane waves with \emph{frequency in the flat and nearly flat band} when the wave amplitude is above a critical value. As a particular case, we consider the sawtooth lattice with Kerr nonlinearity and to represent incoming plane waves, the medium is periodically end-driven.

\begin{figure}[!tbhp]
\includegraphics[angle=0, width=0.9 \linewidth]{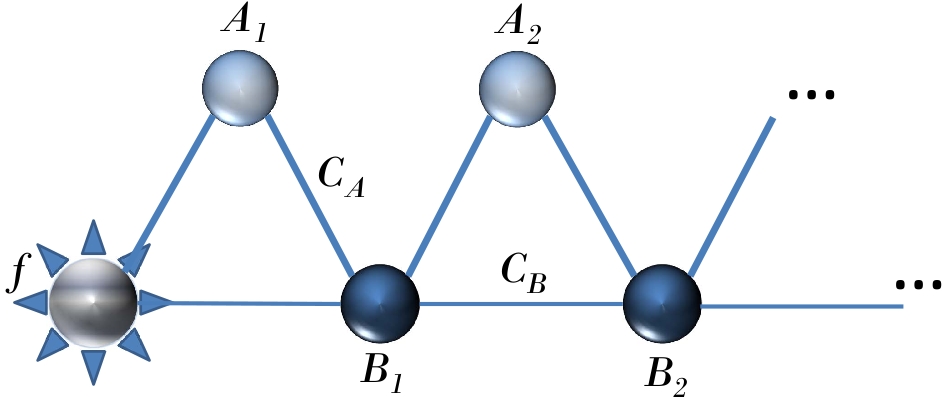}
\caption{Schematic representation of the semi-infinite one-dimensional lattice in the model, driven at the edge by $f$.}
\label{fig1} 
\end{figure}

The sawtooth lattice (portrayed in Fig.\ \ref{fig1}) belongs to a class of quasi-one-dimensional lattices with an accidental flat band, i.e., a flat band that is formed only when a special relation between its coupling coefficients holds \cite{rhim2019classification,maimaiti2017compact}. The presence of an accidental flat band was observed on waveguides arranged on a sawtooth lattice by fine-tuning the inter-site coupling coefficients \cite{weimann2016transport}. 

Sawtooth lattices have been investigated in many diverse physical contexts including photonics \cite{Ji2018,Du2020}, Bose-Hubbard systems \cite{Gremaud2017}, 
attractive Hubbard model \cite{Orso2022}, strongly interacting bosons \cite{Phillips2015}, Bose-Einstein condensates \cite{Wang2020}, and ultracold atoms 
\cite{Zhang2015}. Because completely flat bands of the sawtooth lattice are sensitive to perturbations (e.g., of the coupling coefficients), in the following 
we consider the general case of nearly flat bands, where the characteristics of wave inhibition and supratransmission will be shown to be generic.

%-------------------------------------------------------------------------------
% Figure 1./
%-------------------------------------------------------------------------------

We also study the mechanism behind the surge of transmission phenomenon. In the standard case, it is caused by the generated sequence of gap solitons propagating 
in the medium. Supratransmission in lattices with a flat band was considered in \cite{Motcheyo-Diaz2023}, where generation of gap solitons with zero frequency drive was shown to be possible. %and the effect of on-site potentials on supratransmission was explored \cite{Bountis-Diaz2023}.
However, similar nonlinear modes cannot exist within flat or nearly flat bands. The supratransmission reported herein is therefore different. In fact, we show that in the case of a completely flat band, the surge is due to the instability of a nonlinear edge state with flat band frequency. 

{\em Model equation.} 
The semi-infinite sawtooth lattice is modeled by
\begin{nalign}
    i\dot{A}_n&=C_A\left(\Delta_+B_{n-1}-2A_n\right)+\gamma|A_n|^2A_n,\\
    i\dot{B}_n&=C_A\left(\Delta_+A_n-2B_n\right)+C_B\Delta B_n+\gamma|B_n|^2B_n,
    \label{gov}
\end{nalign}
where the upper dot is the derivative with respect to the time variable $t$ (or the propagation distance in nonlinear optics), $n\in\mathbb{Z}^+$, 
$\Delta_+X_n=X_n+X_{n+1}$, $\Delta X_n =X_{n-1} +X_{n+1} -2X_n$, and $\gamma$ is the nonlinearity coefficient. The lattice is driven through its left boundary 
by setting $B_0 =f(t)$, with $f(t) =F(t) e^{-i\Omega t}$. The amplitude of this harmonic force is turned on adiabatically, e.g., as 
$F(t) =f_0 \left(1-\exp(-t/\tau)\right)$, to avoid initial shock.

\begin{figure}[]
	\includegraphics[angle=0, width=0.9\linewidth]{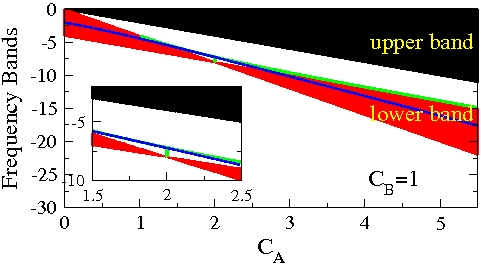}
	\caption{Two allowed bands of the system as a function of the coupling coefficient $C_A$. 
                      Additionally, there is an in-gap eigenfrequency for \emph{finite} lattices shown as green curve (see the text). 
                      Analytical approximation $\hat{\omega}_-$ \eqref{app} is shown in blue. 
                      Inset: Enlargement around $C_A =2 C_B =2$.  
	}
	\label{figband} 
\end{figure}

The linear plane wave of the (infinite) lattice is given by $\{A,B\}_n=\{a,b\}e^{i(kn-\omega t)}$, where
\begin{align}
    \omega_\pm=C_B\left(\cos k-1\right)-2C_A\pm\sqrt{\tilde{\omega}},
    \label{spec}
\end{align}
and $\tilde{\omega}=C_B^2(\cos k -1)^2+2C_A^2(\cos k +1)$, with the corresponding eigenvectors $(a,b)=(C_A(1+e^{-ik}),\omega_\pm+2C_A)$. 
This implies that the lattice has two linear bands at $-2C_A\leq\omega\leq0$ and $-\text{max}(v)\leq\omega\leq-\text{min}(v)$, %$-2(C_A+2C_B)\leq\omega\leq-4C_A$ 
where $v=\{C_A+2C_B,4C_A\}$. The latter %second, i.e., the lower, 
band becomes flat when $C_A=2C_B$ \cite{maimaiti2017compact,rhim2019classification}. We plot the bands in Fig.\ \ref{figband}. Without loss of generality, 
we take $C_B=1$.

%-------------------------------------------------------------------------------
% Figure 3./
%\begin{figure}[tbhp!]
%-------------------------------------------------------------------------------

When the driving frequency $\Omega$ of the external force lies within a linear band, there will be power flow through the lattice even for 
low amplitude driving \cite{khomeriki2004nonlinear,geniet2002energy}. However, note that the flow speed is given by the group velocity 
$\partial\omega/\partial k$. Because $\partial\omega/\partial k\to0$ as %one approaches the flat band condition 
$C_A\to2C_B$, what happens when the driving frequency $\Omega$ is at the flat or in a nearly flat band? This is the main question that we consider in this paper.

{\em Supratransmission.} 
Equations (\ref{gov}) were numerically integrated in time using a sixth order Runge-Kutta scheme \cite{Sarafyan1972} with time-step $h=5\times 10^{-3}$. 
The semi-infinite lattice is truncated into a finite one with $N=256$ unit cells, and dissipation is imposed by adding the terms $-i \Gamma_n A_n$ and 
$-i \Gamma_n B_n$ on the right-hand-side of Eqs.\ (\ref{gov}), respectively, to suppress wave reflection from the right boundary of the 
lattice. The dissipation array $\Gamma_n$ is zero for $n=1, \dots ,N-N_d$, and $\Gamma_n =\Gamma_{max} (n-N+N_d) / N_d$ for 
$n=N-N_d+1, \dots ,N$ ($\Gamma_{max}=5$). Herein, $N_d =32$.
%-------------------------------------------------------------------------------

\begin{figure}[t!]
	\includegraphics[angle=0, width=0.9 \linewidth]{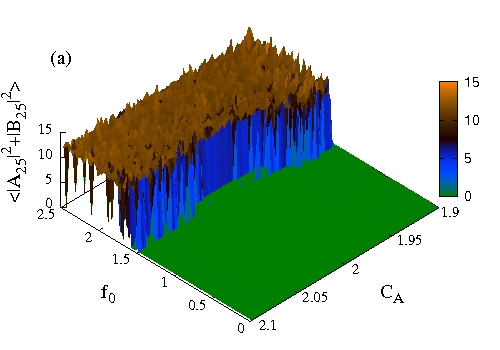}
	\includegraphics[angle=0, width=0.9 \linewidth]{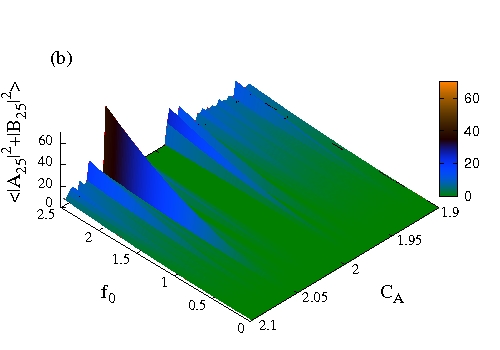}
	\includegraphics[angle=0, width=0.9 \linewidth]{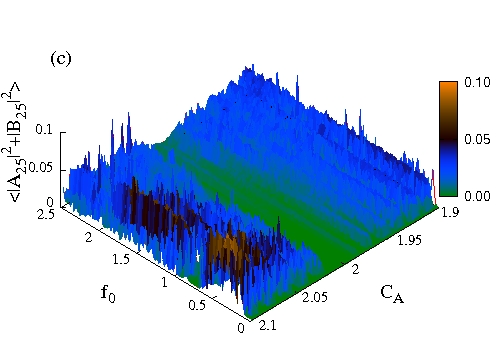}
	\caption{The averaged power at the $25-$th cell, $<|A_{25}|^2 +|B_{25}|^2>$, as
		a function of the driving force amplitude $f_0$ and the coupling 
		coefficient $C_A$ for $C_B =1$, 
		and (a) $\gamma=-2$; (b) $\gamma=0$; and (c) $\gamma=+2$.  
	}
	\label{fig5} 
\end{figure}

Next, we study the flow of power injected by the drive into the array. To show our results, we compute the time-averaged power at a particular site 
$<|A_{25}|^2 +|B_{25}|^2>$, which is far enough from the driving edge. We integrate the governing equation and discard the first $T_{tr} =200,000~T$ 
time-units ($T=2\pi/\Omega$ is the driving period). The integration continues for $T_{av} =50,000~T$ more time-units, where the average power 
is then calculated.  We present it in Fig.\ \ref{fig5} as a function of the driving force amplitude $f_0$ and $C_A$ for several values of $\gamma$. In all cases, 
the driving frequency is $\Omega =-\left(3 C_A +2 C_B\right)$, i.e., in the middle of the lower band.

For the defocusing nonlinearity $\gamma =-2$ the power transmission to the cell at $n =25$ is very low, until a $C_A$-dependent threshold value of 
$f_0 =f_0^{th}$ is reached, where it jumps abruptly to high values of the order of $10$. For $C_A=2$, $f_0^{th}\approx1.668$. Below the threshold, the value 
of the transmitted power is of the order of $10^{-5}$ or less. %, with the lower band being nearly and completely flat. 
For $f_0 > f_0^{th}$, the transmission remains at high levels and nearly constant on average. 
For $\gamma =0$ (linear regime), the transmission increases gradually and nonlinearly with increasing $f_0$. However, different from the defocusing case above, 
the transmission power diminishes at the flat band only. As soon as the lower band is not completely flat, there is a significant power transmission.
For the focusing nonlinearity $\gamma =+2$, a different situation is observed in Fig.\ \ref{fig5}(c), where the transmission is low in the whole 
parameter plane shown with the lowest transmission again around $C_A =2$. 
%the transmission remains low and relatively independent of $f_0$. That transmission in general lowers as $C_A$ approaches $2$, but still remains higher than the corresponding one for the defocusing nonlinearity below $f_0^{th}$. A .  (which are however significantly higher than the corresponding ones for defocusing nonlinearity $\gamma=-2$). The absence of threshold amplitude for power surge in this case will be explained in the next section. 

Thus, the main effect which differentiates these three regimes is the existence of the threshold $f_{th}$ for defocusing nonlinearity which leads to a sharp transition 
from low to high power transmission at that amplitude.

%-------------------------------------------------------------------------------
% Figure 5./
%-------------------------------------------------------------------------------

%-------------------------------------------------------------------------------
% Figure 2./
\begin{figure}[tbhp!]
\includegraphics[width=0.7\linewidth]{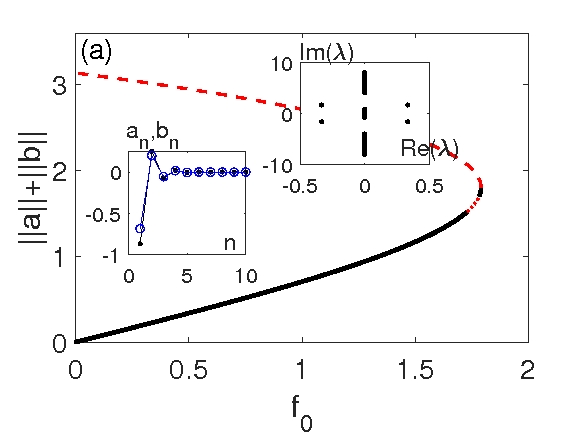}
\includegraphics[width=0.7\linewidth]{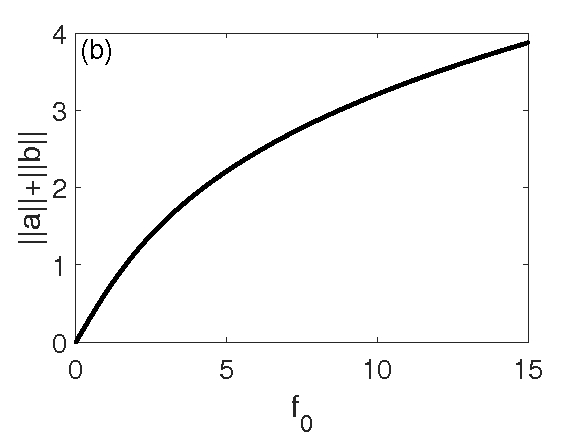}
\caption{Bifurcation diagram of localized solutions of \eqref{stand} as a function of the driving amplitude $f_0$ for (a) $\gamma=-2$ and (b) $\gamma=+2$. 
             Shown is the norm on the vertical axis, where  $||a||=\sqrt{\sum_1^n |{a}_n|^2}$ and  $||b||=\sqrt{\sum_1^n |{b}_n|^2}$.  
             The inset shows the most unstable solution profile on the lower branch and the corresponding linear spectrum 
             in the complex plane. Solid (dashed) lines represent stable (unstable) solutions.}
\label{bif} 
\end{figure}
%-------------------------------------------------------------------------------

{\em Bifurcation of evanescent waves.} 
Supratransmission has been previously shown to be caused by evanescent wave that ceases to exist \cite{khomeriki2004nonlinear,susanto2008boundary,susanto2008calculated} 
in a turning point bifurcation. It is then instructive to consider the standing wave solution of \eqref{gov} with frequency ${\Omega}$. Substituting 
$\{A,B\}_n=\{a,b\}_ne^{-i{\Omega}t}$, we obtain the following coupled algebraic equations
\begin{nalign}
	{\Omega}a_n&=C_A\left(\Delta_+b_{n-1}-2a_n\right)+\gamma a_n^3,\\
	{\Omega}b_n&=C_A\left(\Delta_+a_n-2b_n\right)+C_B\Delta b_n 
	+\gamma b_n^3,
	\label{stand}
\end{nalign}
where $b_0=f_0$. 

We shall look for solution of the system that are localized close to or at the driven end of the sawtooth lattice, while they decay to zero relative far from it. These evanescent modes owe their very existence to the localized eigenmodes associated with the flat band, which are expected to be stable for low $f_0$ (weak nonlinearity).

Once a solution is found, its linear stability is computed through solving the corresponding eigenvalue problem obtained from substituting 
$A_n=(a_n+\hat{a}_ne^{\lambda t})e^{-i{\Omega}t}$, 
$B_n=(b_n+\hat{b}_ne^{\lambda t})e^{-i{\Omega}t}$ into the governing equations \eqref{gov} and 
linearizing about small $\hat{a}_n$ and $\hat{b}_n$. The solution is said to be linearly stable if all the spectrum $\lambda$ has zero real part. 

In Fig.\ \ref{bif}(a) we present a bifurcation diagram of the nonlinear mode as a function of the driving amplitude $f_0$ in the defocusing case ($\gamma=-2$) for 
$C_A=2$. Starting from the lower branch that corresponds to evanescent waves appearing as one ramps up the external drive $f(t)$, we obtain that the corresponding 
solution is stable. As we increase $f_0$ further, there is a critical value where the solution becomes unstable due to a quartet of complex eigenvalues. In the figure, this 
occurs at $f_0^{cr}\approx1.730$, which is in close agreement with the threshold amplitude $f_0^{th}$ reported above.

The small difference between $f_0^{th}$ and $f_0^{cr}$ is due to the strong instabilities which are present in the numerical integration of the dynamical equations
(\ref{gov}) that produce irregular behavior especially when the threshold driving amplitude is approached.

Hence, we obtain that the supratransmission in Fig.\ \ref{fig5}(a) above corresponds to the evanescent wave becoming unstable. This mechanism is different from the 
previously reported cases \cite{geniet2002energy,khomeriki2004nonlinear,susanto2008boundary}, where the threshold amplitude occurs at a turning point. 

%Similar behavior is also expected in other flat band systems, especially those which support localized edge states like the sawtooth lattice.

After instability in the lower branch, as we increase $f_0$ further, we obtain a small portion where the decaying-in-space solution becomes stable again. The localized wave then ceases to exist due to a turning point bifurcation at $f_0^{tp}\approx1.791$. The upper branch corresponds to solutions that are exponentially unstable due to a pair of real eigenvalues. 

We have also calculated nonlinear evanescent waves due to the drive in the focusing case in Fig.\ \ref{bif}(b), where there is no change of stability nor turning point. 
This explains the absence of supratransmission in the focusing case in Fig.\ \ref{fig5}(c).

When the lower band is nearly flat, i.e., $C_A\approx2C_B$, our evanescent wave analysis above cannot be applied because there is no spatially decaying state in the middle of the band. Nevertheless, extreme suppression of power is clearly seen in Fig.\ \ref{fig5}(a,c) when the system is nonlinear. This peculiar phenomenon is caused by nonlinear interaction between the in-band drive and its in-gap excited harmonics. % that are in the We that it is actually a super slow transmission due to a combined effect of (completely or nearly) flat band and nonlinearity. 
The supratransmission in the defocusing nonlinearity case is then caused by the corresponding evanescent modes of the resonant harmonics becoming unstable. This argument will be clearly seen from the time-series analysis below. %, we propose that pseudo-evanescent waves may emerge in the middle of the nearly flat band that originate from the localized edge states supported by the lattice.  These modes also disappear for $f_0$ above a certain threshold, in a way similar to that of the completely flat band case. 

%-------------------------------------------------------------------------------
% Figure S7./
\begin{figure}[tbhp!]
	\includegraphics[angle=0, width=0.95 \linewidth]{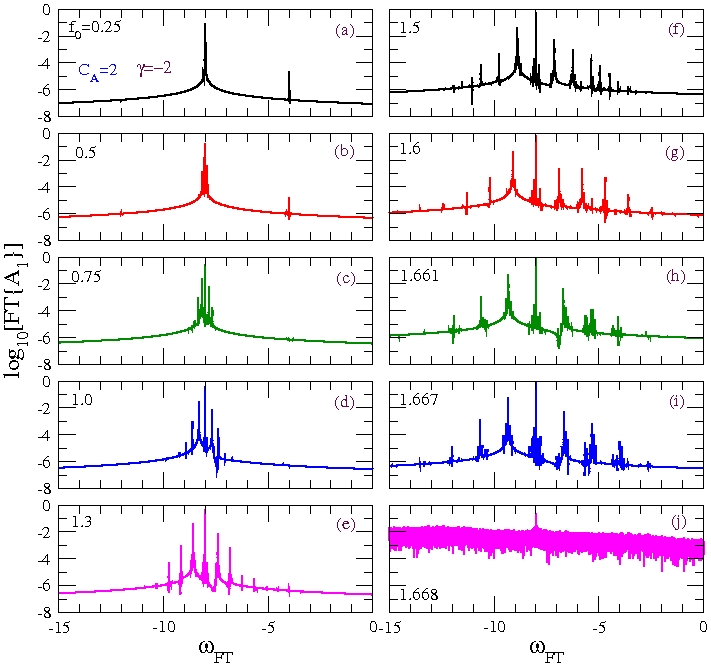}
	\caption{Fourier spectra of $A_1(t)$ for several values of driving amplitudes 
		$f_0$ for $C_B = 1$, $\gamma =-2$, and $C_A =2$, shown in semi-logarithmic plots. %The time-series for each $A_1(t)$ consists of $N_{FT} =2^{19}$ points and span a time-interval of more than $16,000~T$, where $T =2\pi/\Omega$ is the driving period. We discard the first $\tau_{tr} =100,000~T$ time-intervals to eliminate transients. 
	}
	\label{fig-s7} 
\end{figure}
%-------------------------------------------------------------------------------

{\em Time-series analysis.} 
We have performed Fourier spectral analysis of the dynamics of the end-driven lattice. In Fig.\ \ref{fig-s7}, the decimal logarithms of Fourier spectra for $A_1(t)$, $\log_{10}[{FT}\{A_1\}]$, are shown for the complete flat band case $C_A =2$ and several values of the driving amplitude $f_0$. The time series are recorded after integrating the system for $10^5T$, where $T =2\pi/\Omega$ is the driving period, to eliminate transients. The length of each time-series spans a time-interval which corresponds to more than $1.6\times10^4T$; each time series consists of $N_{FT} =2^{19}$ points. The resolution in the Fourier frequencies is therefore $\Delta \omega_{FT} \simeq 2.4\times 10^{-4}$.

For low $f_0$, much lower than the threshold value, the driving frequency is dominant. One may also observe a second (sub)harmonic in the spectra in panels (a,b). The 
frequency, which is at $\omega_{FT}\approx-4$, corresponds to the lower edge of the top band, see Fig.\ \ref{figband}. 

With increasing $f_0$, however, it disappears due to increasing nonlinearity strength that %shifts the lower bound frequency of the upper band away from $-4$ resulting 
results in gradual weakening of the resonance and its subsequent disappearance. At the same time, more subharmonics around the driving frequency appear.

Their separation increases with the increment of $f_0$ and is maintained at an equidistant space for values 
of $f_0$ even very close to the threshold. When $f_0$ exceeds the threshold, the spectrum changes drastically, as it can be observed in panel (j). This change occurs abruptly,
indicating once more that the passage from very low to very high transmission is a real transition. After the transition, the Fourier spectrum becomes essentially continuous 
and rather noisy, while significant power is acquired by all the frequencies in the shown range. Even in this regime, the driving frequency is still visible and dominant.

%%%%%%%%%%%%%%%%%%%%%%%%%%%%%%%%%%%%%%%%%%%%%%%%%%%%%%%%%%%%%%%%%%%%%%%%%%%%%%%%
The equidistant subharmonics appear because the external driving force excites and resonates with an internal edge mode of the semi-infinite system, which can be explained as follows. 

Consider the linear parts of Eqs.\ (\ref{gov}), i.e., set $\gamma=0$ with $f_0=0$, and substitute $A_n=\hat{a}_ne^{-i{\omega}t}$ and $B_n=\hat{b}_ne^{-i{\omega}t}$. We end up with an eigenvalue problem for the eigenfrequency $\omega$ and eigenmodes $(\hat{a}_n,\hat{b}_n)$, that has been solved numerically for the same finite number of unit cells. We plot $\omega$ in Fig.\ \ref{figband} and obtain that in addition to the two continuous bands \eqref{spec}, there is an isolated spectrum shown in green curve.

The eigenfrequency can be observed to get detached from the lower band at $C_A =1$ and persist as an isolated in-gap eigenfrequnecy even for large $C_A$. This frequency is associated with a localized eigenmode having topological origin \cite{Gremaud2017,weimann2016transport}. The inset in Fig.\ \ref{figband} shows it more clearly in a shorter $C_A$ interval. Note that $C_A =2C_B$ is a singular case as the in-gap eigenfrequency merges with the flat band.

The edge eigenmode is highly localized for the parameter value slightly above $C_A =2$. This motivates us to consider the following unit cell at the driven edge  
\begin{nalign}
   i\dot{A}_1&=C_A\left[f+B_1-2A_1\right]+\gamma|A_1|^2A_1,\\
   i\dot{B}_1&=C_A\left(A_1-2B_1\right)+C_B\left[f-2B_1\right]+\gamma|B_1|^2B_1. 
\label{cell}
\end{nalign}  % the matrix notation at the end...
We expect that this system is an approximation of the model \eqref{gov} with an error of $\mathcal{O}\left(1/|\Omega|\right)$ for $|\Omega|\gg1$ 
\cite{susanto2008boundary}. It is standard to obtain that in the linear limit $\gamma=0$, the solution of Eqs. \eqref{cell} involves three frequencies: $\Omega$ and 
\begin{equation}
	\hat{\omega}_\pm=-(2C_A+C_B)\pm\sqrt{C_A^2+C_B^2}.\label{app} 
\end{equation}	
We plot $\hat{\omega}_-$ in Fig.\ \ref{figband}, where good agreement is clearly seen in a relatively wide interval. At $C_A=2$, $\hat{\omega}_-\approx-7.236$ and it agrees well with one of the dominant frequencies in Fig.\ \ref{fig-s7} for $f_0\approx f_0^{th}$. $\hat{\omega}_+\approx-2.764$ approximates the upper band. This shows that the external force $f(t)$ also excites oscillations with frequencies $\hat{\omega}_\pm$. Resonances between $\Omega$ and $\hat{\omega}_-$ through the Kerr nonlinearity then create 
equidistant subharmonic frequencies in Fig.\ \ref{fig-s7}. This also indicates that the system can potentially be a frequency comb generator \cite{fortier201920}.

The distance between subharmonics increases due to increasing nonlinearity strength through $f_0$, since the frequency $\hat{\omega}_-$ depends on the amplitude of the 
waves $A_1$ and $B_1$ when nonlinearities are taken into account.

Features of the spectra in Fig.\ \ref{fig-s7} persist when the band is nearly flat, i.e., $C_A\approx2$. In that case, the excited harmonic frequencies will mostly lie in band gaps that correspond to evanescent waves. These spatially localized modes are responsible for the dispersion suppression in the nonlinear lattice reported in Fig.\ \ref{fig5}, even though frequency of the injected plane wave is in the middle of a linear band. Supratransmission occurs when the modes become unstable. 

\emph{Conclusion.} 
We have shown novel properties of flat and nearly flat band lattices with an injected plane wave. Due to the interaction of their nearly flat band and nonlinearity, wave transmission can be suppressed even when its frequency is inside the band. However, in the defocusing nonlinearity, there is a threshold amplitude for a massive {power surge in the transmission}, i.e., supratransmission. We have shown that this is due to evanescent waves that become unstable. There is no supratransmission for the focusing nonlinearity because such spatially decaying states always exist and are stable. 

We also studied the possibility to employ the system as a frequency comb generator due to nonlinear resonances between the injected wave and the natural edge mode of 
the system. The mode exists because of the topological nature of the lattice.

Our results are particularly relevant to waveguide arrays having the sawtooth configuration, where wave localization due to the presence of an accidental flat band was observed by fine-tuning the inter-site coupling coefficients \cite{weimann2016transport}. Hence, power surge predicted here can be in principle be detected in these systems. Nevertheless, the results will be of interest for researchers in other fields in physics where flat band systems may be realized. 
\\

%%%%%%%%%%%%%%%%%%%%%%%%%%%%%%%%%%%%%%%%%%%%%%%%%%%%%%%%%%%%%%%%%%%%%%%%%%%%%%80
{\em Acknowledgement.} 
This work was supported by a Faculty Start-Up Grant (No.\ 8474000351/FSU-2021-011) by Khalifa University. NL and IK also acknowledged support by Khalifa University through a Competitive Internal Research Awards Grant (No.\ 8474000412/CIRA-2021-064). IK is also grateful for support from KU via the grant No.\ 8474000352/FSU-2021-012.

%%%%%%%%%%%%%%%%%%%%%%%%%%%%%%%%%%%%%%%%%%%%%%%%%%%%%%%%%%%%%%%%%%%%%%%%%%%%%%80
%%\bibliographystyle{unsrt}
%%\bibliography{myfile_V8_Rev}

%%%%%%%%%%%%%%%%%%%%%%%%%%%%%%%%%%%%%%%%%%%%%%%%%%%%%%%%%%%%%%%%%%%%%%%%%%%%%%80
\end{document}